\documentclass[twocolumn,showpacs]{revtex4}
\usepackage{graphicx}%
\usepackage{dcolumn}
\usepackage{amsmath}
\begin{document}
\preprint{}
\title[Short Title ]{Parabolic model with phase-jump coupling\\
}

\author{J. M. S. Lehto}
\email{jaakko.lehto@utu.fi}
\author{K.-A. Suominen}%
\affiliation{Turku Centre for Quantum Physics and Laboratory of Quantum Optics, Department of Physics and Astronomy, University of Turku, FI-20014 Turku, Finland \\
}%

\date{\today}
\begin{abstract}
We study the coherent dynamics of a two-level parabolic model and ways to enhance population transfer and even to obtain complete population inversion in such models. Motivated by the complete population inversion effect of zero-area pulses found in \cite{VasilevVitanov2006}, we consider a scheme where a given coupling function is transformed to a zero-area coupling by performing phase-jump in the middle of the evolution. We also derive a universal formula for the effect of the phase-jump.
\end{abstract}

\maketitle




\section{Introduction}

As far as the idealizations go, the concept of a two-level atom must be both one of the most extreme and most successful ones in the whole field of physics. From the beginning, understanding the coherent dynamics of a two-level quantum system (TLS) interacting with an external field has played an important role in understanding many applications of quantum theory from atomic or molecular collisions \cite{Landau, Zener, Stuckelberg}, magnetic resonance \cite{Rabi1937, RosenZener1932, Majorana}, atom-laser interactions \cite{AllenEberly, Shore2011, vitanov} and   controlling qubits in quantum computation \cite{Stenholm2005}, to name but a few. For many modern applications the aim is to design control pulses that are both very fast and precise and yet robust against any imperfections. It should be also noted that, despite the simple appearance, the number of known soluble models is still very limited \cite{Shore2011, Ishkhanyan2000}. 

In this paper we consider the so-called parabolic model \cite{SuominenIII, Gefen, NakamuraBook}. This model can be applied in situations where the conventional Landau-Zener (LZ) model breaks down, for example in atomic collisions below or at the turning point energy \cite{NakamuraBook} and more recently in inter-band tunnelling near merging Dirac points \cite{Montambaux2012}. The general solution for the model does not exist in a sufficiently compact form \cite{ZhuNakamura1992}. Depending on the parameters, the model can be used to describe different physical situations: interference arising from two level-crossings, a level-glancing situation where the energy levels only touch but do not cross, and transitions happening by Zener tunnelling. In the two latter cases it is not possible to obtain complete population inversion (CPI)  \cite{SuominenIII, LehtoSuominen2012}, which restricts their use in applications. To overcome this deficiency, we consider the possibility of enhancing the population transfer by introducing a phase-jump in the coupling at time $t = 0$. This phase-jump has the effect of transforming the otherwise constant coupling into a zero-area coupling. 

The amount of excitation in TLS resulting from an applied resonant pulse depends only on the total pulse area and not on the details, such as the form or the amplitude, of the pulse \cite{AllenEberly, McCallHahn}. This result is known as the area theorem and it is, among other applications, the basis of the use of the so-called $\pi$ and $\pi/2$ pulses in NMR and quantum computation. It also follows from this theorem that the pulse with zero total area is of a self-cancelling nature and the system is returned to the initial state despite the transient excitation it experiences while the coupling is on. What actually happens with non-resonant zero-area pulses is quite surprising as one might usually expect that the excitation is most efficient with a resonant field. However, as was  found relatively recently \cite{VasilevVitanov2006}, the system can exhibit CPI in the strong coupling region under quite general requirements and, unlike the resonant case, it is robust. 

This CPI effect has been demonstrated for both smooth zero area pulses \cite{VasilevVitanov2006, LehtoKAS2016} and pulses with a sudden jump in phase to make them antisymmetric in time \cite{VitanovPhaseJump}, while the detuning is kept constant. The mechanism behind CPI derives from the extreme non-adiabatic behavior of the system in the strong coupling limit. This also allows one to obtain an area theorem type of approximative formula for the transition probability \cite{VasilevVitanov2006}. 

The other motivation for our work, besides finding ways to enhance population transfer for certain models, is to understand the CPI effect of \cite{VasilevVitanov2006} better. By studying the propagators in the two bases formed either by the bare states or the eigenstates of the Hamiltonian, we can identify the contribution of the phase-jump to the dynamics. This allows us to give an alternative derivation of the approximate formula for the transition probability given in \cite{VasilevVitanov2006} or \cite{VitanovPhaseJump} and address its universal character.

This paper is organized as follows. In section \ref{sec:Formalism} we present the basic equations and definitions. The section \ref{sec:GenDiscussionDyn} concentrates on the effect of the phase-jump coupling on the dynamics on a general level, while in section \ref{sec:ParabolicModel} we introduce the parabolic model and derive the approximative formulas for the transition probability. The discussion in \ref{sec:Conclusions} concludes the presentation.


\section{Mathematical formalism}
\label{sec:Formalism}

\subsection{The Schr\"{o}dinger equation}

Since we are discussing the coherent dynamics of two-level and certain symmetry considerations play an important role in our discussion, the Schr\"{o}dinger equation governing the evolution of the system is best given in the propagator form (in units where $\hbar = 1$)
\begin{equation}
\dot{\imath} \partial_{t} U(t, t_{0}) = H(t)U(t, t_{0}), 
\label{eqn:SchEq}
\end{equation}
where $U(t, t_{0})$ is the unitary matrix propagating the arbitrary initial state vector of the system at time $t_{0}$ to time $t$, 
\begin{equation}
\psi(t) = U(t, t_{0})\psi(t_{0}),
\label{eqn:PropagatorDef}
\end{equation}
which implies that the proper initial condition in (\ref{eqn:SchEq}) is $U(t_{0}, t_{0}) = 1$ for any $t_{0}$. 

\subsection{The diabatic and adiabatic bases}

The time-dependence of the Hamiltonian is due to external fields interacting with the two-state system and we start by specifying the Hamiltonian in the so-called diabatic (or bare) basis whose basis states refer to the two stationary states in the absence of the interaction. It is given by the Hermitian matrix (we use the subscript D to denote this basis) 
\begin{eqnarray}
H_{D}(t) &= \vec{H}(t)\cdot\vec{\sigma} \\ \nonumber
		 &= \left( \begin{array}{lr}
			\alpha (t) & V(t)e^{-\dot{\imath}\phi} \\
			V(t)e^{\dot{\imath}\phi} & -\alpha (t)
					\end{array}\right), 
\label{eqn:hbare}		 		
\end{eqnarray}
where the field vector is defined as $\vec{H}(t) = \left(V(t)\cos(\phi), V(t)\sin(\phi), \alpha(t) \right)^{T}$ and the components of $\vec{\sigma}$ are the Pauli matrices, as usual. The real functions $V(t)$ are refered to as diabatic energy levels and diabatic coupling, respectively. Usually, the Hamiltonian can be given in the real-symmetric form (we can fix $\phi = 0$) and this also our starting point. However, the zero-area coupling models studied in this paper can be understood as involving a jump of $\pi$ in the phase of the coupling, so it is useful to keep the notation general, although for us it suffices to consider the phase as piecewise constant. Furthermore, the state vector in this basis is denoted by $\psi_{D}(t) = \left( C_{e}(t), C_{g}(t) \right)^{T}$ where $C_{e}$ and $C_{g}$ are the amplitudes of the diabatic basis states. 

Another important representation of the dynamics is obtained by deploying the instantaneous eigenstates of the system as the basis states. The transformation to this so-called adiabatic basis (denoted by subscript A) is given by $\psi_{A}(t) = R(t)\psi_{D}(t)$, where the matrix
\begin{equation}
R(t) =  \left( \begin{array}{lr}
			\cos(\frac{\theta(t)}{2}) & -\sin(\frac{\theta(t)}{2})e^{-\dot{\imath}\phi} \\
			\sin(\frac{\theta(t)}{2})e^{\dot{\imath}\phi} & \cos(\frac{\theta(t)}{2})
					\end{array}\right),
\label{eqn:basischange}					
\end{equation}
has the eigenstates as its columns and it is defined $\tan(\theta(t)) = V(t)/\alpha(t)$. In this basis the Hamiltonian is given by 
\begin{equation}
H_{A}(t)  = \left( \begin{array}{lr}
			 E_{+}(t) & -\dot{\imath}\gamma(t)e^{-\dot{\imath}\phi} \\
			 \dot{\imath}\gamma(t)e^{\dot{\imath}\phi} & E_{-}(t)
					\end{array}\right), 
\label{eqn:hadiabatic}		 		
\end{equation}
where the eigenvalues are $E_{\pm} = \pm \sqrt{\alpha^{2}(t) + \vert V(t)\vert^{2}}$ and the adiabatic coupling is 
\begin{eqnarray}
\gamma(t) &= \frac{\alpha(t)\dot{V}(t) - \dot{\alpha}(t)V(t)}{2\left(\alpha^{2}(t) + V^{2}(t) \right)} \nonumber \\
		  &\equiv \frac{\dot{\theta}(t)}{2}, 
\label{eqn:acoupl}
\end{eqnarray}
where the overhead dot stands for time-derivative. The nice thing is, that these two bases usually coincide in the initial and final times when we have $\vert \alpha(t) \vert \gg \vert V(t)\vert$. Furthermore, as we usually take the intial state to be $\vert C_{g}(t_{0}) \vert = 1$ the transition probability in the final time $t_{f}$, $P_{D} \equiv \vert C_{e}(t_{f})\vert^{2}$, is easily given also in the adiabatic basis by either $P_{A} = P_{D}$ or $1 - P_{D} $.

In the following we use various expressions for matrices in different bases interchangeably and the subscripts $D$ and $A$ are used throughout to highlight which one of the representations is used to express a particular matrix. 

\subsection{Zero-area pulses and phase-jump coupling}

A resonantly excited two-level system undergoes Rabi flopping between the two states. Explicitly, if the system is initially in the ground state, the excited state probability is given by 
\begin{equation}
P = \sin^{2}\left[\frac{A(t, t_{0})}{2}\right],
\label{eqn:Presonance}
\end{equation}
where the pulse area is defined as 
\begin{equation}
A(t, t_{0}) = 2\int_{t_{0}}^{t}V(x)\mathrm{dx}.
\label{eqn:PulseArea}
\end{equation}
This result is known as the area theorem, which states that the final population resulting from the pulse depends solely on the total pulse area \cite{AllenEberly, McCallHahn}. 
It also implies that with a vanishing total area of the coupling, the system always returns to the initial state. This symmetry is generally lost when the field is off-resonant, although it still holds when both the diabatic energy levels and coupling are odd functions of time \cite{VitanovKnight1995}. 

However, the constant non-zero detuning and an antisymmetric diabatic coupling can result, quite unexpectedly, even in CPI and to be robust against parameter variations \cite{VasilevVitanov2006}. This happens in the strong coupling limit and with detuning larger then the bandwidth of the pulse. The CPI effect can be explained by the delta-function type of behavior of the adiabatic coupling near time $t = 0$ and actually applying the area theorem in the adiabatic basis. This effect is present both for smooth antisymmetric pulses (for which $V(0) = 0$) \cite{VasilevVitanov2006} and for pulses which reverse their sign non-continuously at $t = 0$ \cite{VitanovPhaseJump}. This latter can be also interpreted as a phase jump from $\phi = 0$ to $\phi = \pi$ at $t = 0$. 

\section{General discussion of the dynamics}
\label{sec:GenDiscussionDyn}

On the basis of the results in the previous section, in this paper, we want to consider the models with zero-area coupling but which also have time-dependent diabatic energy levels. In particular, from the basis of the results in \cite{VitanovKnight1995}, we restrict the reference model to be of the form 
\begin{eqnarray}
\alpha_{ref}(-t) &= \alpha_{ref}(t)\\
V_{ref}(-t) &= V_{ref}(t).
\label{eqn:RefModelForm}
\end{eqnarray}

With the reference model given, we can write its zero-area variant with the phase-jump coupling as 
\begin{eqnarray}
\tilde{\alpha}(t) &= \alpha_{ref}(t)\\
\tilde{V}(t) &= \left(2h(-t) - 1\right)V_{ref}(t),
\label{eqn:ModelForm}
\end{eqnarray}
where $h(x)$ is the step function having the value zero for negative and unity for positive arguments. The corresponding zero-area model Hamiltonian is also denoted by $\tilde{H}(t)$. This Hamiltonian then coincides with the reference Hamiltonian for negative times and therefore the propagator from $t_{0} = -\infty$ to $t = 0$ is the same for both models. In the following, we therefore split the total propagator into two parts,
\begin{equation}
S = U(\infty, 0)U(0, -\infty) 
\label{eqn:scattering}
\end{equation}
and study its connection to the propagator of the variant model,
\begin{eqnarray}
\tilde{S} &= \tilde{U}(\infty, 0)\tilde{U}(0, -\infty) \\
&=  \tilde{U}(\infty, 0)U(0, -\infty).
\end{eqnarray}

Before discussing the phase-jump dynamics in detail in different bases, we note the general connection of the propagators given in diabatic and adiabatic bases. If we assume that the equation (\ref{eqn:PropagatorDef}) is given in the diabatic basis to begin with, then by using equation (\ref{eqn:basischange}), it reads in the adiabatic basis
\begin{equation}
\psi_{A}(t) = R(t)U_{D}(t, t_{0})R^{\dagger}(t_{0})\psi_{A}(t_{0}),
\end{equation}
which leads to the connection
\begin{equation}
U_{A}(t, t_{0}) = R(t)U_{D}(t, t_{0})R^{\dagger}(t_{0}),
\label{eqn:Connection}
\end{equation}
which is not generally a similarity transformation because the $R$ matrices are evaluated at different points.

\subsubsection{Description of the phase-jump dynamics in the diabatic basis}

Because the operation $\sigma_{z}M\sigma_{z}$ simply changes the sign of the off-diagonal components of any $2\times2$ matrix $M$, the connection between the reference and zero-area Hamiltonians at positive times is simply given by 
\begin{equation}
\tilde{H}_{D}(t) = \sigma_{z}H_{D}(t)\sigma_{z}, \quad t \geq 0
\end{equation}
and it follows from (\ref{eqn:SchEq}) that the similar equation holds for the propagators
\begin{equation}
\tilde{U}_{D}(t, 0) = \sigma_{z}U_{D}(t, 0)\sigma_{z}, \quad t \geq 0.
\label{eqn:tildeU}
\end{equation}
This just means that if the total evolution for the reference model is given in the diabatic basis by equation (\ref{eqn:scattering}), then we simply have
\begin{equation}
\tilde{S}_{D} = \sigma_{z}U_{D}(\infty, 0)\sigma_{z}U_{D}(0, -\infty).
\end{equation}

For later purposes, it should be also noted that any propagator that is pure phase evolution, 
\begin{equation}
U_{\varphi} =  \left( \begin{array}{lr}
			e^{\dot{\imath}\varphi} & 0 \\
			0 & e^{-\dot{\imath}\varphi}
					\end{array}\right), 
\label{eqn:phaseevolution}
\end{equation}
commutes with this transformation and is therefore left invariant $U_{\varphi} = \sigma_{z}U_{\varphi}\sigma_{z}$.

\subsubsection{Appearance of the phase-jump dynamics in the adiabatic basis}

In the adiabatic basis, the situation is not so simply described due to the extreme non-adiabatic behavior at $t = 0$. It can be understood considering the evolution in the Bloch sphere picture \cite{Shore2011}. Initially, $\theta(-\infty) = 0$ and the field vector starts at the north pole and moves in the x-z plane $(\phi = 0)$. If the system is initially in one of the basis states and the dynamics is adiabatic, the state follows the field vector or its antipodal point. Moreover, if $V(t) \gg \alpha(t)$ as $t\rightarrow 0$ then the field vector tends to the equator of the Bloch sphere and $\theta = \pi/2$. What happens to the eigenstates $\chi_{\pm}(t)$ there, is that when $t < 0$ we have, from (\ref{eqn:basischange}),
\begin{equation}
\chi_{+}(t) = \frac{1}{\sqrt{2}} \left( \begin{array}{c} 1 \\
									  1 \end{array} \right) , \qquad \chi_{-}(t) = \frac{1}{\sqrt{2}} \left( \begin{array}{c} -1 \\
										   1 \end{array} \right),
\end{equation}
whereas right after the phase-jump at $t = 0$ we have $\phi = \pi$ and 
\begin{equation}
\tilde{\chi}_{+}(t) = \frac{1}{\sqrt{2}} \left( \begin{array}{c} 1 \\
									  -1 \end{array} \right) , \qquad \tilde{\chi}_{-}(t) = \frac{1}{\sqrt{2}} \left( \begin{array}{c} 1 \\
										   1 \end{array} \right),
\end{equation}
so that the eigenstates change their character. So if the system is initially, say, in the ground state $\chi_{-}$, after the phase-jump it is in state $-\tilde{\chi}_{+}$. The rest of the evolution is again adiabatic and therefore the population of the system transfers completely to the excited state. This CPI, of course, happens only if the evolution is strictly adiabatic except for the phase-jump. 
%

\section{The parabolic model}
\label{sec:ParabolicModel}

\begin{figure}
\includegraphics[scale=0.65]{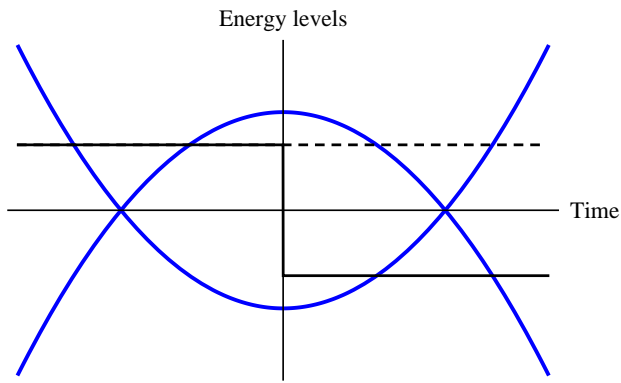} 
\caption{(Colour online) The schematics of the parabolic model are plotted. Thick blue lines are for the diabatic energy levels, while the solid black line is for the diabatic coupling with a phase jump. The black dashed line is for the coupling of the reference model.}
\label{fig:levels}
\end{figure}

We now introduce our reference model which is the parabolic model \cite{SuominenIII, Gefen, NakamuraBook}
\begin{eqnarray}
\alpha(t) &= a t^{2} - c \\
V(t) &= b,
\end{eqnarray}
where $a$ and $b$ are positive real parameters and the value of $c$ can be either positive, zero or negative. These model functions are plotted in figure \ref{fig:levels}. As there is freedom in eliminating one of the parameters, we do this by fixing $a \equiv 1$. This model is discussed at length elsewhere (see, for example, \cite{SuominenIII}) and we only review its basic properties shortly. The value of the parameter $c$ divides the model into three cases: double-crossing ($c > 0$), level-glancing ($c = 0$) or tunneling case ($c < 0$). In each case the transition probability $P_{D} = P_{A} \equiv P$ and goes to zero in both the weak coupling/sudden ($b \rightarrow0$) and the strong coupling/adiabatic limits ($b \rightarrow \infty$).


In the double-crossing case, the diabatic levels cross at two times $t_{c} = \pm \sqrt{c}$. As there is ambiguity to which energy level the system takes  between the crossings, these trajectories interfere and there are oscillations in the final populations of the system. Also, CPI is obtainable with suitable parameters, $P_{max} = 1$. 

In the glancing case, the diabatic levels only touch at the origin and it can not be understood as a Landau-Zener type process. Furthermore, it seems that CPI can not be obtained, $P_{max}$ being only little over one half \cite{SuominenIII, LehtoSuominen2012}.

\begin{figure}[hb]
\includegraphics[scale=0.6]{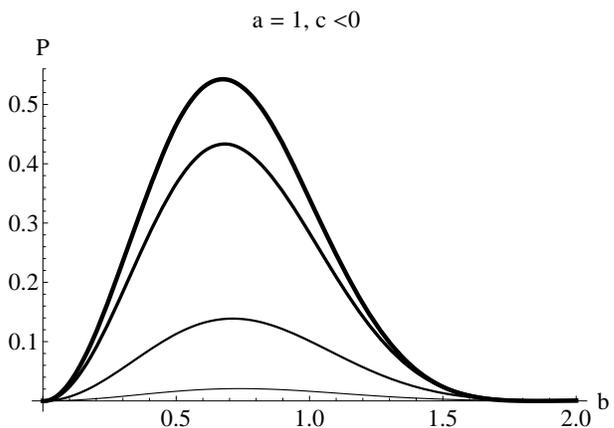} 
\caption{Transition probability for the reference parabolic model for the case $c \leq 0$. The different values are, from thickest to thinnest, $c = 0$, $c = -0.1$, $c = -0.5$ and $c = -1$. The transition probability in the tunnelling case is quickly suppressed. }
\label{fig:RefTunnellingPics}
\end{figure}

When $c < 0$ we have no crossings but the transitions are happening only by Zener tunnelling and they are strongly suppressed; see figure \ref{fig:RefTunnellingPics}.

\subsection{Independent crossing approximation for the reference model}

An important special case amenable for analytical description is the parameter region where the two crossings are well separated. This happens when $c$ is large and it is well motivated physically to consider the two crossings as uncorrelated events \cite{NakamuraBook}. Then one can linearize the diabatic energy levels in the vicinity of the crossings and apply the Landau-Zener theory. As this involves two subsequent LZ events, the scattering matrix which can be given for a single LZ process, is best given in the adiabatic basis \cite{Kazantsev, TorosovVitanovSuperpositionsLZ, SuominenIII},
\begin{equation}
S_{A} =  \left( \begin{array}{lr}
			\sqrt{1 - R^{2}}e^{\dot{\imath}\phi_{S}} & -R \\
			R & \sqrt{1 - R^{2}}e^{-\dot{\imath}\phi_{S}}
					\end{array}\right),
\label{eqn:LZpropagatorA}					
\end{equation}
where $R = \exp\left(-\pi\Lambda/2\right)$ is the amplitude of the LZ transition and 
\begin{equation}
\phi_{S} = \frac{\pi}{4} + \frac{\Lambda}{2}\ln \left(\frac{\Lambda}{2 e}\right) + \mathrm{arg}\left[\Gamma\left(1 - \dot{\imath}\Lambda/2\right)\right]
\end{equation}
is the Stokes phase. The effective LZ parameter is obtained from the linearization as $\Lambda = 2\sqrt{ac}$. Now, the full parabolic model evolution is composed as 
\begin{equation}
S_{A}^{tot} = S_{A,2}U_{\varphi_{dyn}}S_{A, 1},
\label{eqn:SAtot}
\end{equation}
where the $S_{A, i}$, $i = 1, 2$ are the two LZ events and the accumulation of the dynamical phase between the crossings is different for different adiabatic levels,
\begin{equation}
\varphi_{dyn} = 2\int_{0}^{\sqrt{c/a}}\mathrm{ds}\sqrt{(a s^{2} - c)^{2} + b^{2}},
\end{equation}
and must be taken account with matrix (\ref{eqn:phaseevolution}). Also the first scattering matrix $S_{A, 1}$ is just given by the equation (\ref{eqn:LZpropagatorA}) but the second has to be modified slightly, 
\begin{equation}
S_{A, 2} = \sigma_{z} S_{A, 1} \sigma_{z} .
\label{eqn:SA2}
\end{equation}
This looks like the equation (\ref{eqn:tildeU}) but one must note that we are now working in the adiabatic basis. Indeed, this modification is actually due to the fact that the non-adiabatic coupling of the parabolic model is an odd function of time. By dividing the evolution into two LZ processes means that we have to additionally take this structure into account in this approximation and this is just what equation (\ref{eqn:SA2}) accomplishes \cite{SuominenIII}. 

\begin{figure}[h]
\includegraphics[scale=0.7]{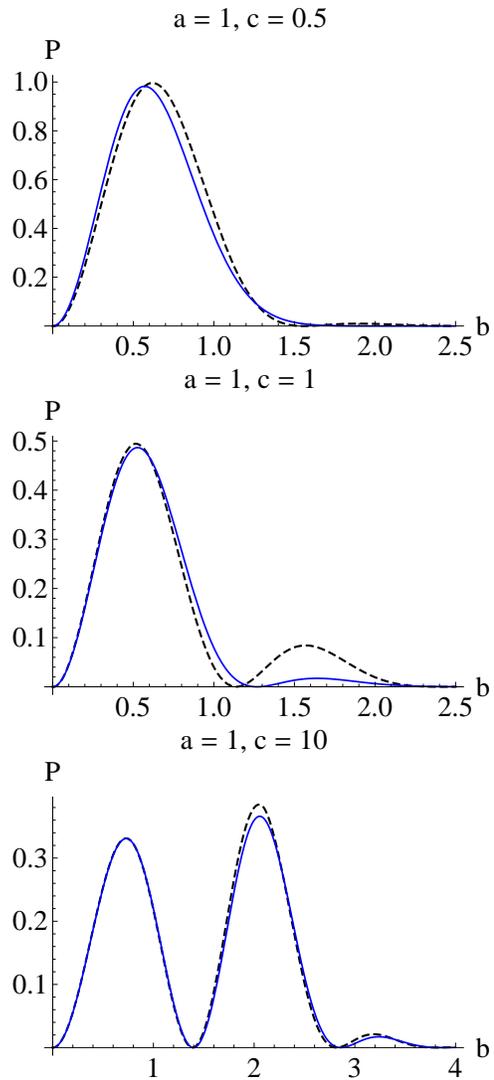} 
\caption{(Colour online) The transition probability for the double-crossing case of the reference parabolic model is drawn with dashed black line for the cases $c = 0.5$, $c = 1$ and $c = 10$. The independent crossing approximation of equation (\ref{eqn:refP}) is drawn with blue line.}
\label{fig:RefDCPics}
\end{figure}

Calculating the equation (\ref{eqn:SAtot}), we obtain the well-known expression for the transition probability \cite{SuominenIII}
\begin{equation}
P = 4R^{2}(1 - R^{2})\sin^{2}\left(\varphi_{dyn} +  \phi_{S}\right),
\label{eqn:refP}
\end{equation}
and figure \ref{fig:RefDCPics} demonstrates that this formula is a good approximation for $P$, regardless of the value of the coupling $b$, when $c \gg 1$ and the crossings are well separated. For smaller positive values of $c$, equation (\ref{eqn:refP}) is mostly useful with smaller couplings, although it does give  the correct limiting value $P \approx 0$ for strong coupling.

In the following, we calculate the total scattering matrix for the phase-jump model to obtain the corresponding $\tilde{P}$.

\subsection{Independent crossing approximation with the phase-jump coupling}

In this case, it is useful to consider the scattering matrix separately on time regions $t < 0$ and $t \geq 0$, so we denote $S_{tot} = S_{+}S_{-}$. Also the evolution related to the dynamical phase between the crossings is correspondingly divided to contributions from the negative and positive times and denoted $U_{\varphi_{dyn}} \equiv U_{+}U_{-}$. Because the reversal in the sign of the diabatic coupling was so easily understood in the diabatic basis, as discussed in section \ref{sec:GenDiscussionDyn}, we must transform the propagators in the previous subsection to the diabatic basis. This is done by applying the equation (\ref{eqn:Connection}). For $t < 0$, noting that $R(-\infty)$ is identity, 
\begin{equation}
S_{D, -} = R^{\dagger}(0)U_{-}S_{A}.
\end{equation}
Similarly, for $t \geq 0$ we have
\begin{equation}
S_{D, +} = R^{\dagger}(\infty)\sigma_{z}S_{A}\sigma_{z}U_{+}R(0),
\end{equation}
where $R^{\dagger}(\infty)$ is in the case of parabolic model at most some phase evolution that does not give a non-trivial contribution to the dynamics and can be left out. Combining these directly  as $S_{D, +}S_{D, -}$ would give the total propagator of the reference parabolic model in the diabatic basis. It should be noted that in this case the rotation matrices (\ref{eqn:basischange}) evaluated at $t = 0$ give identity and this just leads to the transition probability (\ref{eqn:refP}), as it should. 

Now, when the phase-jump in the diabatic coupling is taken into account with equation (\ref{eqn:tildeU}), we get the full propagator in the diabatic basis as
\begin{eqnarray}
\tilde{S}_{D}^{tot} &= \sigma_{z}S_{D, +}\sigma_{z}S_{D,- } \\
&= \sigma_{z}^{2}S_{A}\sigma_{z}U_{+}R(0)\sigma_{z}R^{\dagger}(0)U_{-}S_{A}\\
&= S_{A}\sigma_{z}U_{+}R(0)\sigma_{z}R^{\dagger}(0)U_{-}S_{A}.
\label{eqn:tildeSD}
\end{eqnarray}
Now the off-diagonal element $\left(S_{D}^{tot}\right)_{12} $ is real, so the transition probability is obtained just by squaring this, 
\begin{eqnarray}
\tilde{P} &= \lbrace  (2R^{2} -1)\sin\left[\theta(0)\right] \nonumber \\ 
&+ 2\sqrt{1 - R^{2}} R \cos\left[\theta(0)\right]\cos\left[\varphi_{dyn} + \phi_{S}\right]     \rbrace^{2},
\label{eqn:tildeP}
\end{eqnarray}
where $\theta(0) = \arctan \left(-\frac{b}{c}\right)$, so that now
\begin{equation}
\sin\left[\theta(0)\right] = \frac{b}{\sqrt{c^{2} + b^{2}}} , \qquad \cos\left[\theta(0)\right] = \frac{-c}{\sqrt{c^{2} + b^{2}}}.
\end{equation}
It is seen that when $\vert c \vert \gg b$, the first line in (\ref{eqn:tildeP}) vanishes and the transition probability $\tilde{P}$ is the same as for the original parabolic model but with the argument of the oscillatory part shifted by $\pi/2$. In general there is a contribution from both of the terms and the expression is more complicated as shown in figure \ref{fig:DCplots}.

\subsection{Universal formula for the dynamics of phase-jump coupling}

Mathematically, the main difference between the reference and phase-jump models comes from the fact that there is a $\sigma_{z}$ matrix between the rotation matrices evaluated at $t = 0$ as mentioned in the previous subsection. This gives 
\begin{equation}
R(0)\sigma_{z}R^{\dagger}(0) = \left( \begin{array}{lr}
			\cos(\theta(0)) & \sin(\theta(0)) \\
			\sin(\theta(0)) & -\cos(\theta(0))
					\end{array}\right),
\end{equation}
instead of the identity that is obtained in the reference case. It is easy see from equation (\ref{eqn:tildeSD}) that when we are in the adiabatic parameter region (for parabolic model this is the strong coupling region $b \gg 1$), this is also the only non-trivial contribution in the dynamics as the matrices $S_{A}$ are diagonal and contribute only to phase evolution. 

Therefore, we see that related to the phase-jump of the coupling, there is a universal approximate transition probability, in the sense that it depends only on  the value $\theta(0)$ calculated at the time of the jump and not on other details of the model. This is given by 
\begin{eqnarray}
\tilde{P}_{0} &= \sin^{2}[\theta(0)] \\
&= \frac{V^{2}(0)}{V^{2}(0) + \alpha^{2}(0)},
\label{eqn:universalP}
\end{eqnarray}
This is non-oscillatory and tends monotonously to unity as $\left(V(0)/\alpha(0)\right)^{2}$ increases, obtaining the value one half always at $\vert \alpha(0) \vert = V(0)$. 

\begin{figure}[h]
\includegraphics[scale=0.7]{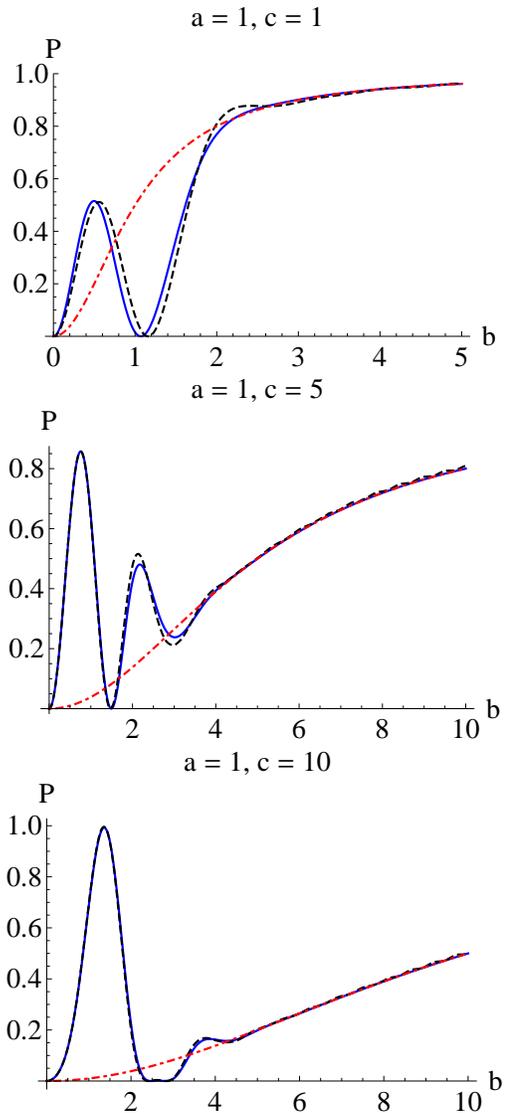} 
\caption{(Colour online) Here we have plotted the transition probability $\tilde{P}$ for the zero-area parabolic model in the double-crossing case for three different values of $c$. The solid blue line is the independent crossing approximation, red dot-dashed line is the universal approximation of equation (\ref{eqn:universalPparabolic}) while the black dashed line is the numerical result.}
\label{fig:DCplots}
\end{figure}

\begin{figure}[h]
\includegraphics[scale=0.6]{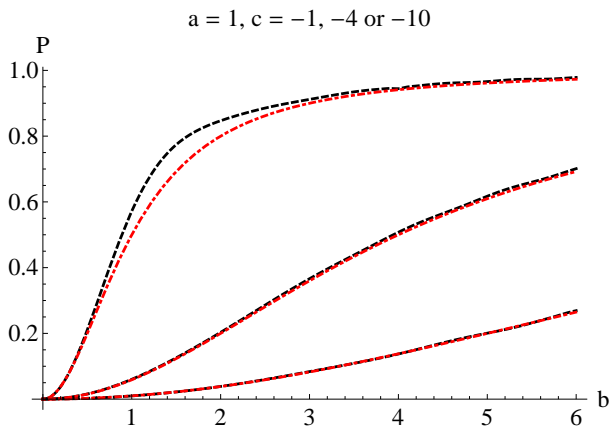} 
\caption{(Colour online) Here we have plotted the transition probability $\tilde{P}$ for the zero-area parabolic model for negative $c$. The numerical results are plotted with black dashed and the red dot-dashed line is the universal approximation of equation (\ref{eqn:universalPparabolic}). The values of the parameter $c$ are as follows: $c = -1$ (uppermost curves), $c = -4$ (middle) and $c = -10$ (lower).}
\label{fig:TunnelingPlots}
\end{figure}

For the parabolic case we have 
\begin{equation}
\tilde{P}_{0} = \frac{b^{2}}{b^{2} + c^{2}},
\label{eqn:universalPparabolic}
\end{equation}
which, if valid, would assign the same value for the double-crossing and tunnelling cases with the same value of $\vert c \vert$. The applicability of this formula is shown in figures \ref{fig:DCplots} and \ref{fig:TunnelingPlots}.  

\section{Conclusions}
\label{sec:Conclusions}

\begin{figure}[h]
\includegraphics[scale=0.6]{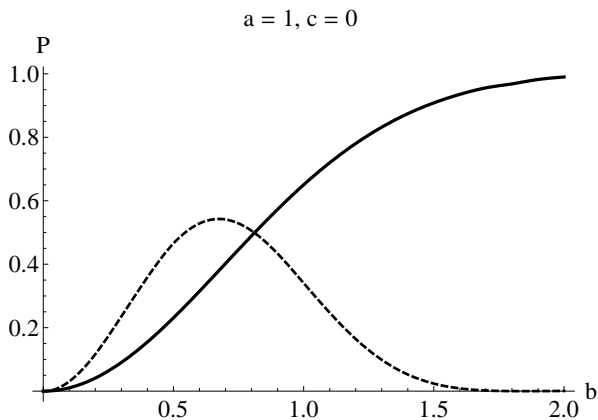} 
\caption{In this plot, the influence of the phase-jump coupling is demonstrated for the parabolic level-glancing case. The transition probability of the  reference model is plotted as the dashed line, whereas the corresponding case with the phase-jump coupling is given by the solid line.}
\label{fig:ComparisonGlancing}
\end{figure}

We have discussed generally the dynamics of two-level models with zero-area coupling in the idealized case where the coupling is instantly flipped. Our analysis shows clearly that it is possible to affect greatly to the transition probability by rapidly changing the phase of the coupling from $\phi = 0$ to $\phi = \pi$ and that this enhancement of the transitions happens not just for the constant non-zero detuning but also when we drive the diabatic energy levels in some fashion. This can be particularly important in experimental situations where the system parameters at one's disposal are restricted somehow and the transitions do not happen efficiently enough. The parabolic level-glancing case in figure \ref{fig:ComparisonGlancing} exemplifies this fact. There, with the phase-jump coupling one obtains CPI with relatively small coupling, in contrast with the maximum transition probability of about one half of the reference model. This same fact is even more dramatic for the case of Zener tunnelling, where the transition probability is normally highly suppressed and very close to zero. However, with a flip in the sign of the coupling, $\tilde{P}$ can obtain significant values and even CPI for strong-enough coupling, regardless of the height of the tunnelling barrier ($c \ll 0$), as can be seen from figure \ref{fig:TunnelingPlots}.

From figure \ref{fig:DCplots} and from equation (\ref{eqn:tildeP}) it is clear that the dynamics exhibits two different regions for the double-crossing case. The final transition probability is oscillating function of the coupling for small and moderate values of the coupling, whereas for larger values it tends monotonously to unity. Both of these characteristics are well described by the independent crossing approximation (\ref{eqn:tildeP}). Indeed, this formula works remarkably well for every value of the coupling even for quite small separation between the crossings and the approximation seems to be applicable even  better for the phase-jump model than for the original double-crossing model (see, for example, the case with value $c = 1$ in figures \ref{fig:RefDCPics} and \ref{fig:DCplots}).

We also derived an approximative formula, equation (\ref{eqn:universalP}) which describes how the transition probability $\tilde{P}$ tends to unity when the coupling increases. For the parabolic model, with large enough $\vert c \vert$, it also coincides with the true $\tilde{P}$ in the parameter region of $b$ where the oscillations are over and the behavior of $\tilde{P}$ is monotonic. In particular, the figure \ref{fig:TunnelingPlots} shows that it is an excellent approximation for the tunnelling case. When $ \vert c \vert$ is closer to zero the universal formula still gives qualitatively correct, although neither of the approximations is really suited to that region. Figure \ref{fig:ComparisonGlancing} is an example of such case and it is seen that CPI is obtained also there from quite small coupling value onwards.

The formula (\ref{eqn:universalP}) can also be found in \cite{VasilevVitanov2006} and \cite{VitanovPhaseJump} where it has been obtained either by applying the area theorem in the adiabatic basis in the strong coupling region or as a part of an analytic solution of a certain soluble model. Our method, instead, was to divide the evolution into negative and positive time periods and to consider the general expressions for the propagators in different bases. In this way, one sees that it is a general feature of the phase-jump coupling and it is most clearly seen when the evolution of the reference model would otherwise be trivial, i.e., in either adiabatic or tunnelling regions. The universal character of the formula should be also highlighted. The approximation $\tilde{P}_{0}$ depends only on the values of the diabatic energy and coupling in one point of time. For example, if we consider more general superparabolic models \cite{LehtoSuominen2012, Lehto2013} given by  $\alpha_{ref}(t) = t^{2n} - c$ where $n = 1, 2, 4, \ldots$, this approximation is the same for all of the models in this class. Indeed, the numerical simulations show that the tunnelling solutions, which are well described by $\tilde{P}_{0}$ coincide for different superparabolic models.  

In conclusion, we have found that the zero-area couplings are a very useful tool to greatly improve transitions between states of a TLS and that when the  diabatic energies of the system are also driven in a suitable fashion, in our case with parabolic time-dependency, the CPI effect can be obtained even with quite small couplings.

\end{document}